\documentclass[aps, superscriptaddress, showpacs, nofootinbib, prd,longbibliography, citecolor=blue,colorlinks=true]{revtex4-2}
\usepackage[utf8]{inputenc}
\usepackage[T1]{fontenc}
\usepackage{ae,aecompl} 
\usepackage{graphicx}
\graphicspath{{Images/}} 
\usepackage{amsmath}
\usepackage{color}
\usepackage{xcolor}
\usepackage{amssymb}
\usepackage{orcidlink}
\usepackage{latexsym}
\usepackage{wasysym}
\usepackage{psfrag}
\usepackage{ifthen}
\usepackage{hyperref}
\usepackage{longtable}
\usepackage{float}
\usepackage[utf8]{inputenc}
\usepackage{lineno}
\usepackage{units}
\usepackage{multirow}
\usepackage{ulem}

\begin{abstract}
This document presents an overview of the design, implementation, and expected performance of the Advanced Virgo Plus (AdV+) upgrades in view of the O5 observing run. Following the experience gained during the O4 commissioning and operations, the Virgo Collaboration has revised the upgrade strategy to address limitations associated with marginally stable recycling cavities. The O5 upgrade program combines elements from the original AdV+ Phase II project with new design solutions, including the implementation of stable recycling cavities, a major modification to the central interferometer layout, and a comprehensive renewal of critical subsystems. The planned upgrades are organized in two steps, targeting progressive improvements in operational stability, noise reduction, and detector sensitivity. Key developments include new vacuum infrastructures, suspensions, mirrors, optical configurations, quantum noise reduction systems, and high-power laser technologies. The resulting configuration is expected to significantly enhance the interferometer performance, enabling a substantial increase in astrophysical reach and scientific return during O5.
\end{abstract}

\begin{document}

\title{Advanced Virgo Plus for O5 \\
Design Report Overview}


\author{F. Acernese}
\affiliation{Dipartimento di Farmacia, Universit\`a di Salerno, I-84084 Fisciano, Salerno, Italy}
\affiliation{INFN, Sezione di Napoli, I-80126 Napoli, Italy}
\author{A. Agapito\, \orcidlink{0009-0005-9004-3163}}
\affiliation{Centre de Physique Th\'eorique, Aix-Marseille Universit\'e, Campus de Luminy, 163 Av. de Luminy, 13009 Marseille, France}
\author{D. Agarwal\, \orcidlink{0000-0002-8735-5554}}
\affiliation{Universit\'e catholique de Louvain, B-1348 Louvain-la-Neuve, Belgium}
\author{I.-L. Ahrend}
\affiliation{Universit\'e Paris Cit\'e, CNRS, Astroparticule et Cosmologie, F-75013 Paris, France}
\author{L. Aiello\, \orcidlink{0000-0003-2771-8816}}
\affiliation{Universit\`a di Roma Tor Vergata, I-00133 Roma, Italy}
\affiliation{INFN, Sezione di Roma Tor Vergata, I-00133 Roma, Italy}
\author{A. Ain\, \orcidlink{0000-0003-4534-4619}}
\affiliation{Universiteit Antwerpen, 2000 Antwerpen, Belgium}
\author{S. Albanesi\, \orcidlink{0000-0001-7345-4415}}
\affiliation{Theoretisch-Physikalisches Institut, Friedrich-Schiller-Universit\"at Jena, D-07743 Jena, Germany}
\affiliation{INFN Sezione di Torino, I-10125 Torino, Italy}
\author{W. Ali}
\affiliation{INFN, Sezione di Genova, I-16146 Genova, Italy}
\affiliation{Dipartimento di Fisica, Universit\`a degli Studi di Genova, I-16146 Genova, Italy}
\author{C. All\'en\'e}
\affiliation{Univ. Savoie Mont Blanc, CNRS, Laboratoire d'Annecy de Physique des Particules - IN2P3, F-74000 Annecy, France}
\author{A. Allocca\, \orcidlink{0000-0002-5288-1351}}
\affiliation{Universit\`a di Napoli ``Federico II'', I-80126 Napoli, Italy}
\affiliation{INFN, Sezione di Napoli, I-80126 Napoli, Italy}
% ...

\collaboration{The Virgo Collaboration}



\maketitle

\section{Advanced Virgo Plus: Phase II-O5}
Advanced Virgo Plus (AdV+) is an upgrade of Advanced Virgo implemented in two stages: Phase I and Phase II~\cite{PhaseOneReport}.

The main goal of Phase I was to reduce the limitation coming from quantum noise by adding a frequency dependent vacuum squeezed sources and by implementing the remaining parts of the Advanced Virgo upgrade~\cite{AdV_2015} namely the installation of the signal recycling mirror and the increase of the power from 25W to 40W. 
Other improvements aimed at reducing the effect of various technical noises. The bulk of Phase I has been built and installed between the fall of 2019 and the spring of 2021. The commissioning of the main interferometer could start in January 2021 and the commissioning of the quantum noise reduction system, including the new 300 m long filter cavity, started in May 2021. 

\begin{figure}[ht!]
\centering
\includegraphics[scale=0.32]{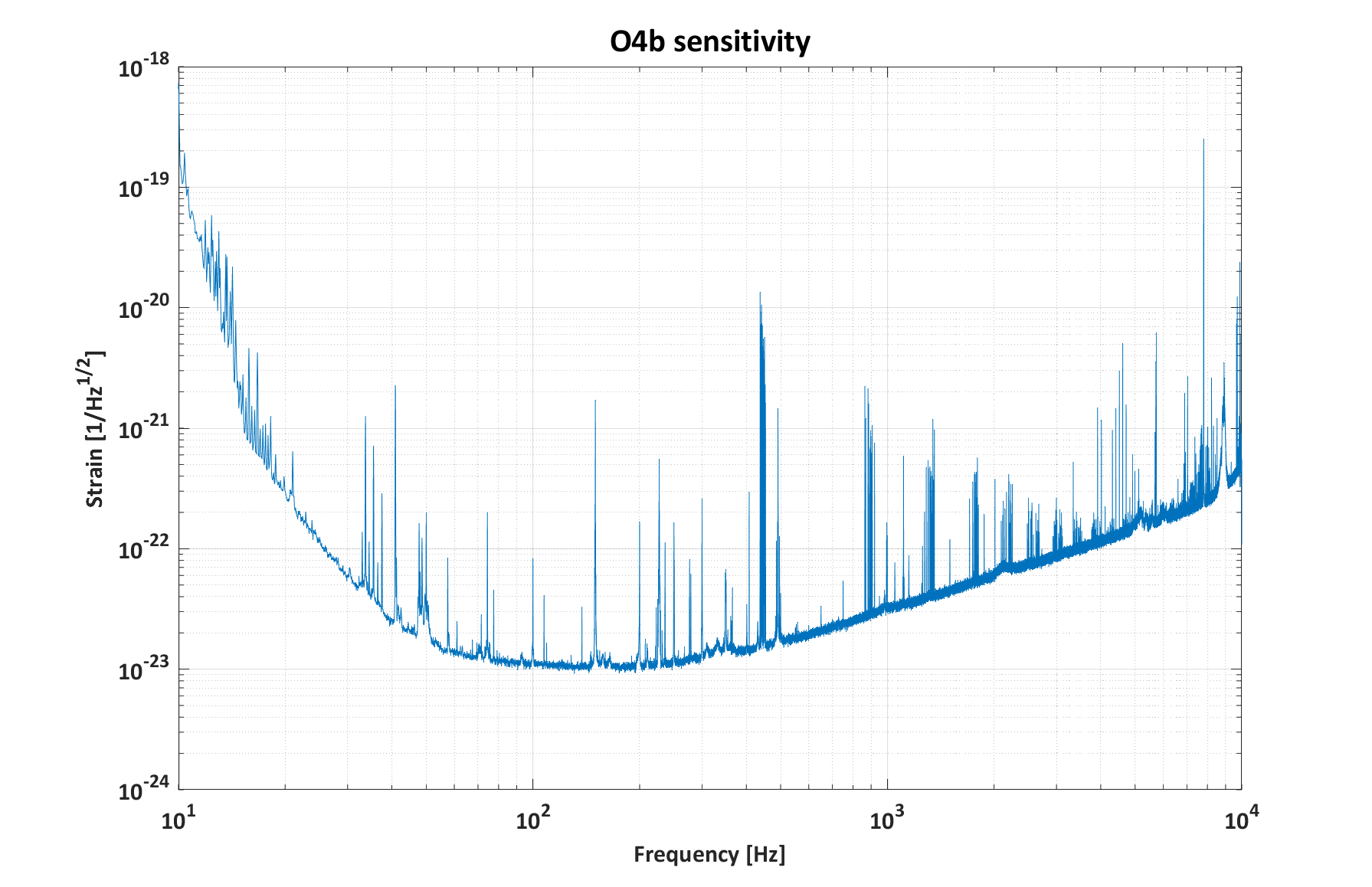}
\caption{Typical sensitivity of the Virgo detector during the run O4b}.
\label{fig:O4bsens}
\end{figure}

These tuning phases have highlighted the challenges of operating marginally stable recycling cavities in a dual-recycled interferometer. To ensure stable operation and minimize the noise, it was necessary to reduce the input power to 17 W—thus mitigating thermal effects—and to misalign the signal recycling mirror by 2 $\mu$rad.
These adaptations have allowed Virgo to join the two LIGO detectors in the O4 observing run in May 2024, with a sensitivity (shown in Figure~\ref{fig:O4bsens}), expressed as the average range for the detection of the coalescence of a binary neutron star system, of about 55 Mpc.

Given that the issues encountered during Phase I commissioning get worse as the power stored in the interferometer is increased, which makes reaching the design value of the input power for Advanced Virgo Plus Phase II (80 W in O5)~\cite{PhaseTwoReport} not realistic, the Virgo Collaboration has decided to implement stable recycling cavities after O4~\cite{StabCavReport, SSClayout, SSCDes}.

The addition of the stable cavities to the AdV+ Phase II Project~\cite{PhaseTwoReport} has, inevitably, triggered an overall review of the scope of the project in view of O5. In particular, it was proposed to postpone the installation of the large end test masses (and of the other related modifications) to post-O5 upgrade. Furthermore, a survey was conducted among the subsystem managers to identify the upgrades highlighted during the commissioning phase in preparation for O4.
The Virgo upgrade project for O5 includes components from the previous AdV+ Phase II, elements required for the installation of stable recycling cavities, and items identified during the O4 commissioning phase.

The main difference with respect to Phase I will be the insertion of two off-axis telescopes in the recycling cavities in order to make them stable (see Figure~\ref{fig:OptConf}). The optical configuration has been designed so that the higher-order mode fields have a different resonance condition than the carrier field. This is in contrast to the marginally stable cavities currently implemented in Virgo, where the higher-order modes are practically co-resonant with the carrier. 

As a consequence, while the 3-km arm cavities will remain unchanged, the layout of the central interferometer will be significantly modified. The old vacuum chambers, hosting the power and signal recycling mirrors and the suspended injection and detection benches, together with their seismic isolation systems, will be removed through two openings to be cut in the roof of the central building. Eight new vacuum chambers and their links~\cite{SSCvacta} will be deployed to accommodate the recycling cavity mirrors and the new suspended optical benches.

Other changes related to the vacuum system concern the pumping stations that will be added along the arms to reduce the pressure in the tubes. In addition, several ion pumps will be installed in the central area to reduce the noise coming from the pumping system.

Given the deep change of the optical configuration in the central interferometer, the benches hosting the telescopes used to inject the light into the interferometer and to extract it at the output port have to be completely revisited~\cite{daumas2026presentation}. At each of the input and output ports, one of the currently existing bench will be replaced by two suspended benches, comprising baffles rigidly connected to the benches themselves. One of these new benches will be sharing its suspension system with the new recycling mirror. At the output port this bench will also host Faraday Isolators with larger aperture, as well as an ultra fast shutter required to protect downstream optics from lasers flashes in the new interferometer configuration.

The optics and benches will be suspended through a new compact seismic isolation system, based on the AdV Multi-SAS~\cite{Heijningen_2019}/ET pathfinder building blocks, and equipped with on-purpose payloads.

The suspended optics of the recycling cavities and the vacuum links connecting the new vacuum chambers will be provided with baffles, for absorbing diffused light. Inside the vacuum chambers of the input test masses, a new type of baffle~\cite{InstBaff} able, not only to absorb the light scattered around the mirror, but also to monitor it, will be installed. To this purpose, these baffles will be equipped with several photodiodes and their readout electronics.

In parallel with the change of the configuration of the recycling cavities, the laser power injected into the interferometer will be increased from the present 17 W to 25-80 W. To reach this goal the power at the output of the pre-mode-cleaner will have to be increased up to 115 W and the injection system throughput will be pushed up to 70\%. A new laser amplifier system similar to the one used in LIGO will be installed.

The larger laser power will require improving the thermal compensation system. Ring heaters~\cite{Nardecchia_2023} and central heaters~\cite{Accadia_2013} will be installed on the power and signal recycling mirrors, in order to fine-tune the radii of curvature of these optics and improve control over the Gouy phase of the cavities. The telescopes for the Hartman wave-front sensors will be changed, following the new optical layout of the central interferometer. Finally, yet importantly, the CO2 laser projector will be upgraded to improve the quality of the heating pattern and the amount of power available for the compensation.

The O5 upgrade will also be the occasion to upgrade part of the control system workstations and the oldest part of the timing distribution system. Moreover, there is a plan to renew a large fraction of the injection and laser electronics dating from initial Virgo. Finally, the obsolescent electronics controlling the Superattenuators will be replaced.

Several upgrades concern the electronics and the controls of the filter cavity of the frequency dependent squeezing system. At the same time, the auxiliary green laser system, used in the lock acquisition sequence of the interferometer, will be adapted to the new optical layout of the recycling cavities and upgraded to improve its robustness. Finally yet importantly, the calibration system will be upgraded to reduce the systematic errors.

\section{Configuration of AdV+ for O5}
\begin{figure}[ht!]
\centering
\includegraphics[scale=0.65]{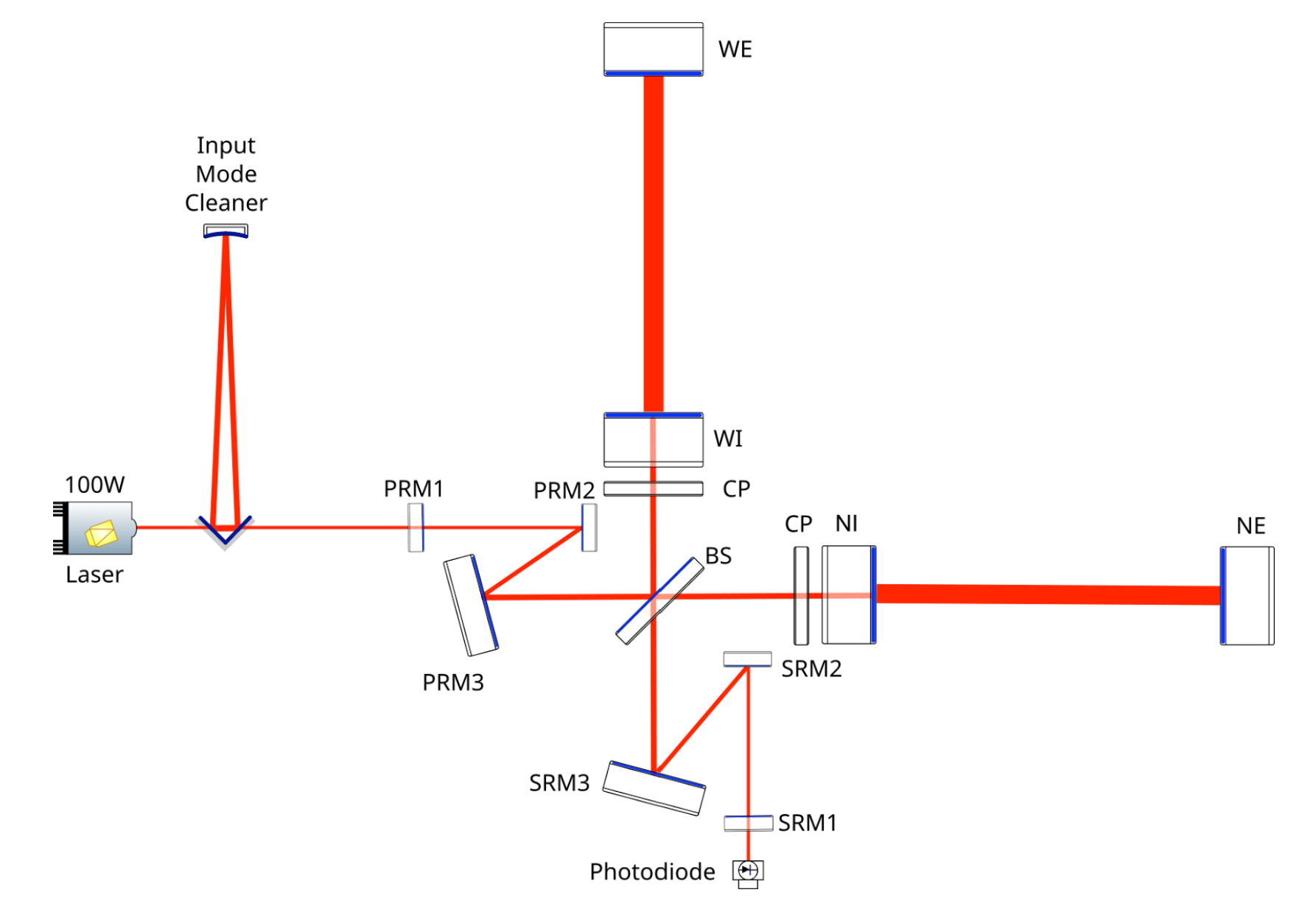}
\caption{Configuration of Advanced Virgo + in view of O5. The main difference compared to Phase I will be the folding mirrors in the recycling cavities. Modifications will be made also to the vacuum system.}
\label{fig:OptConf}
\end{figure}

The optical configuration of Advanced Virgo+ for O5 is shown in Figure~\ref{fig:OptConf}.

The main parameters of AdV+ in O5 are summarized in Table~\ref{tab:virgo_comparison}. For comparison, the parameters of Virgo at the beginning of O4b and during O4c are shown.

\begin{table}[htbp]
\centering
\small
\setlength{\tabcolsep}{6pt}
\renewcommand{\arraystretch}{1.2}
\begin{tabular}{|l|c|c|}
\hline
\textbf{Parameter} & \textbf{Virgo during O4b/c} & \textbf{AdV+ O5} \\
\hline
\multicolumn{3}{|l|}{\textbf{Laser \& Injection}} \\
\hline
Laser power & 100 W & 100--150 W \\
Power into the interferometer & 17 W & 25--80 W \\
Power in the arm cavities & 96 kW & 120--380 kW \\
Input mode-cleaner length & 143 m & 141.2 m \\
Input mode-cleaner finesse & 1000 & 1000 \\
Modulation frequencies & 6, 8, 56 MHz & 6, 15, 27, 78 MHz \\
\hline
\multicolumn{3}{|l|}{\textbf{Interferometer Optical Configuration}} \\
\hline
Arm cavities length & 3 km & 3 km \\
Arm cavities finesse & 450 & 450 \\
Power recycling gain & 39 & 39 \\
Signal recycling & Yes (tilted by 2 $\mu$rad) & Yes \\
RoC input mirrors & 1420 m & 1420 m \\
RoC end mirrors & 1683 m & 1683 m \\
g-factor & 0.87 & 0.87 \\
Beam size on mirrors & 49 mm / 58 mm & 49 mm / 58 mm \\
PRC/SRC degeneracy & Degenerate & Non-degenerate \\
\hline
\multicolumn{3}{|l|}{\textbf{Mirrors}} \\
\hline
Beam splitter & 55 cm $\times$ 6.5 cm, 34 kg, T=50\% & 55 cm $\times$ 6.5 cm, 34 kg, T=50\% \\
Input mirrors & 35 cm $\times$ 20 cm, 42 kg, T=1.4\% & 35 cm $\times$ 20 cm, 42 kg, T=1.4\% \\
End mirrors & 35 cm $\times$ 20 cm, 42 kg, T=5 ppm & 35 cm $\times$ 20 cm, 42 kg, T=5 ppm \\
Power recycling mirror & 35 cm $\times$ 10 cm, 21 kg, T=5\% & 15 cm $\times$ 10 cm, 4 kg, T=5\% \\
Signal recycling mirror & 35 cm $\times$ 10 cm, 21 kg, T=40\% & 15 cm $\times$ 10 cm, 4 kg, T=40\% \\
\hline
\multicolumn{3}{|l|}{\textbf{Coatings}} \\
\hline
Coating materials & Ti:Ta$_2$O$_5$ / SiO$_2$ & Ti:Ta$_2$O$_5$ / SiO$_2$ \\
\hline
\multicolumn{3}{|l|}{\textbf{Detection}} \\
\hline
Output mode-cleaner & Single cavity, Finesse = 1000 & Single cavity, Finesse = 1000 \\
Detection losses & 15\% & 6\% \\
Photodiode quantum efficiency & 99\% & 99\% \\
\hline
\multicolumn{3}{|l|}{\textbf{Suspensions}} \\
\hline
Mirror suspensions & Monolithic, fused silica fibers & Monolithic, fused silica fibers \\
Vibration isolation & Super-attenuators & Super-attenuators + multi-SAS \\
\hline
\multicolumn{3}{|l|}{\textbf{Quantum Noise Reduction System}} \\
\hline
Injected squeezing & 8 dB & 11 dB \\
Filter cavity & L = 285 m, Finesse = 10000 & L = 285 m, Finesse = 10000 \\
Injection losses & 11\% & $<$ 10\% \\
Phase noise & 120 mrad & 40 mrad \\
\hline
\multicolumn{3}{|l|}{\textbf{Vacuum}} \\
\hline
Residual pressure in arm cavities & $4\times10^{-9}$ mbar & $2.5\times10^{-9}$ mbar \\
Residual pressure in TM towers & $2\times10^{-8}$ mbar & $1\times10^{-8}$ mbar \\
\hline
\end{tabular}
\caption{Main parameters of AdV+ in O5 compared to those of Virgo at the beginning of O4b. The coatings that will be used in O5 will be decided at the end of the coating R\&D.}
\label{tab:virgo_comparison}
\end{table}

\section{Sensitivity of AdV+ in O5}
\begin{figure}[ht!]
\centering
\includegraphics[scale=0.3]{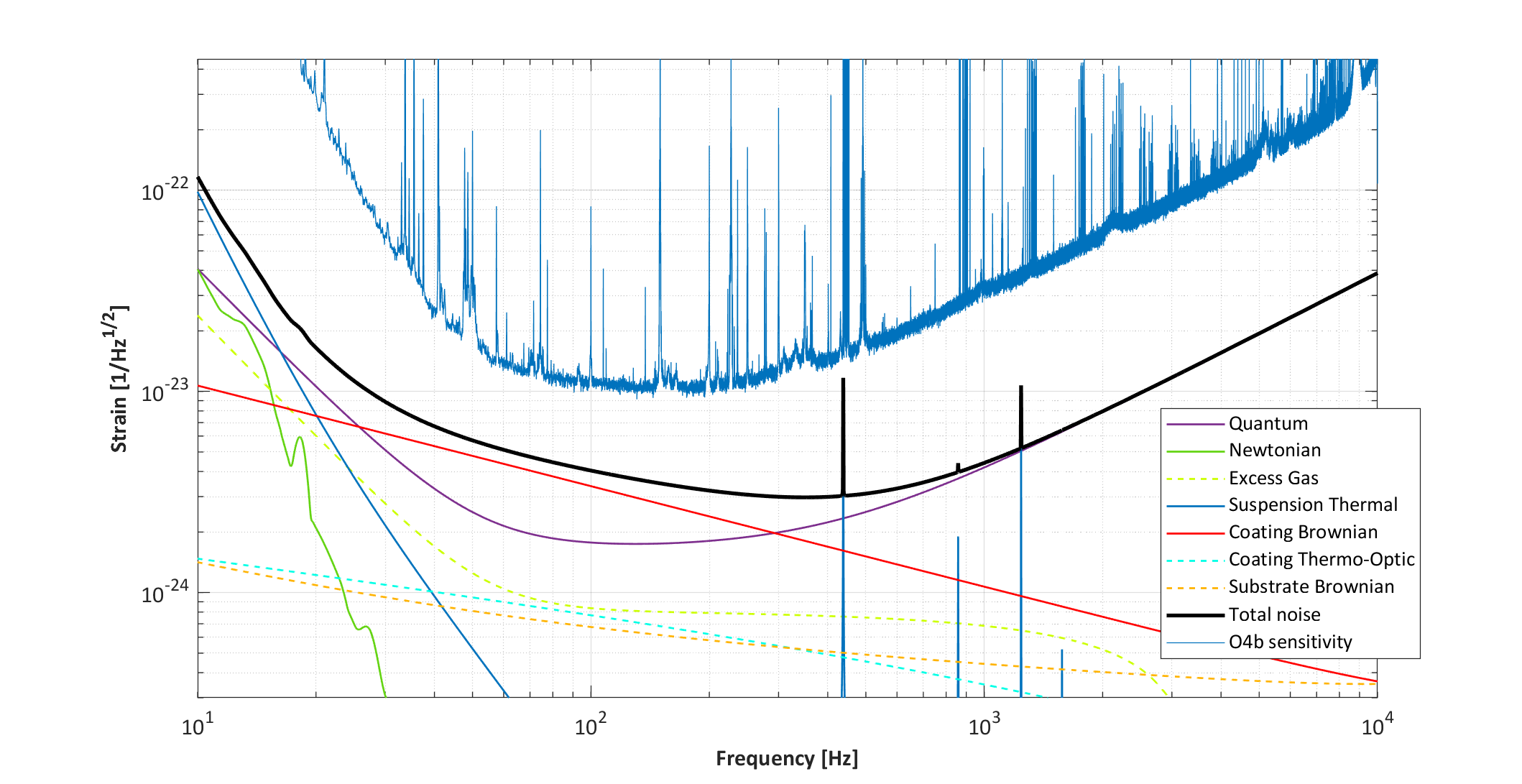}
\caption{Best sensitivity curve for AdV+ in O5. For comparison the sensitivity of AdV during O4b is shown.}
\label{fig:O4bDesign}
\end{figure}

The best sensitivity that AdV+ will be able to reach in O5 is shown in Figure~\ref{fig:O4bDesign}, computed using the Matlab version of Gwinc~\cite{matgwinc}.

The main noises limiting the sensitivity of AdV+ in O5 will be the following:
\begin{itemize}
    \item \textbf{Quantum noise}
The amplitude of this noise is driven by the laser power injected into the interferometer (25 to 80 W), the details of the optical configuration (parameters of the Fabry-Perot cavity, the power recycling mirror and the signal recycling mirror) and the performances of the frequency dependent squeezing. For the latter the sensitivity curve shown in Figure~\ref{fig:O4bDesign} assumes that the optical losses will be such as to allow a reduction of quantum noise by 6 dB over the full interferometer bandwidth.

  \item \textbf{Mirror thermal noise}
This noise is dominated by the coating Brownian noise. The amplitude of this noise is shown in Figure~\ref{fig:O4bDesign} and assumes a reduction of the coating thermal noise by a factor of 70\% compared to the coatings used for Advanced Virgo \cite{CoatTherm}.

  \item \textbf{Pendulum thermal noise}
The pendulum thermal noise is due to the mechanical losses in the payload. It is evaluated from the known losses in the fused silica fibers, the geometry of the wires used to suspend the mirror and the marionette, and the measured marionette mechanical losses.

  \item \textbf{Newtonian noise}
This noise is estimated using the best available model that implements the following parameters: seismic spectrum, seismic speed, and distance from test mass to ground.

  \item \textbf{Technical noises}
This noise is mainly due to the combination of control noise, scattered light, actuators noise and environmental disturbances. Figure~\ref{fig:O4bDesign} assumes that all these technical noise will be reduced below the fundamental noises described above.
\end{itemize}

The sensitivity shown in Figure~\ref{fig:O4bDesign} represents the best theoretically achievable performance, assuming all parameters are pushed to their maximum values. However, due to budgetary and scheduling constraints, reaching this ideal sensitivity is practically impossible. To balance performance goals with available resources and observing time, the upgrade path has been divided into two stages.

Stage 1 includes the installation of stable cavities, together with the new vacuum system, suspensions, mirrors, and suspended benches, as well as other essential upgrades required to ensure reliable interferometer operation during O5 (including new input test masses, the laser source, and control electronics). It also incorporates improvements specifically designed to enhance the detector’s sensitivity. Stage 2 comprises further upgrades aimed at achieving additional sensitivity gains, including new end test masses, a high-power laser source, and improved adaptive optics.

\begin{figure}[ht!]
\centering
\includegraphics[scale=0.4]{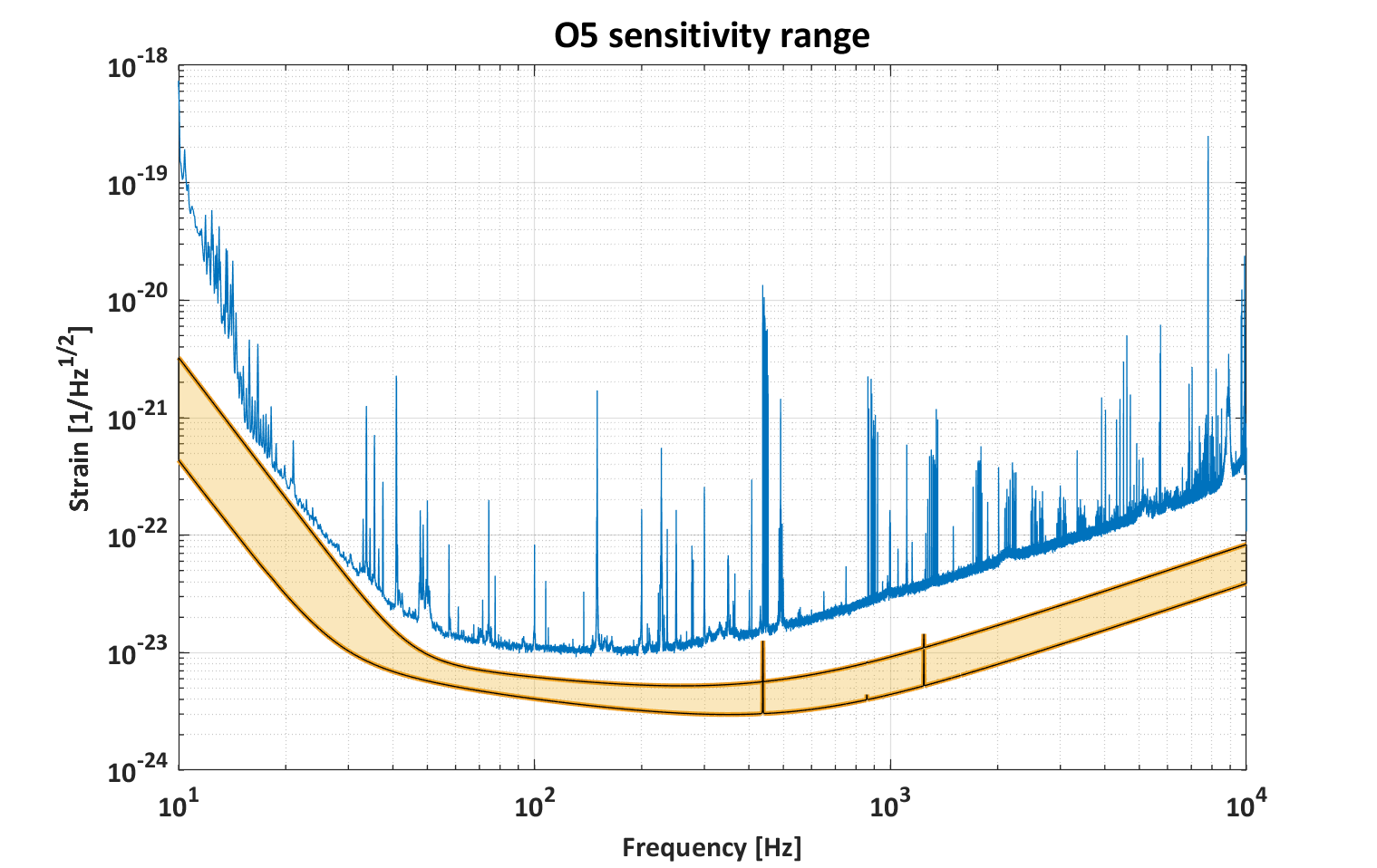}
\caption{Range of sensitivities expected for AdV+ O5 (orange band). Also shown is the sensitivity measured by AdV during O4b in blue.}
\label{fig:Step1Step2}
\end{figure}

Following the framework of the upgrade stages, the expected, and more conservative estimate (computed again using Gwinc~\cite{matgwinc}), of the sensitivity of AdV+ in O5 is shown in Figure~\ref{fig:Step1Step2}. This corresponds to the orange band delimited by the worst Stage 1 and best Stage 2 sensitivities. For comparison the sensitivity measured in O4b is reported.

At high frequency the difference comes from more conservative hypotheses on the laser power injected into the interferometer (from 80 W to 25 W) and on the quantum noise reduction due to the frequency dependent squeezing (from 6 dB to 4.5 dB - as measured in LIGO~\cite{LIGOQND}). At intermediate frequencies, the difference comes from the coating thermal noise. The more conservative curve assumes that the coating materials used during O5 will have the same mechanical losses as the one used for AdV. Finally, at low frequency the difference comes from a more conservative hypothesis on the residual technical noises. With these more conservative assumptions, the binary neutron star range will range between 90 Mpc and 160 Mpc. In the most optimistic case, it will reach 180 Mpc. 

The list of parameters used to evaluate the sensitivities achievable during AdV+ O5 are given in Table~\ref{tab:virgo_o5_steps}. For comparison, the table includes the O4b working parameters. The ASCII files containing the sensitivity curves (both amplitude and power spectral densities) of Figure~\ref{fig:Step1Step2} are publicly available at: \url{https://tds.virgo-gw.eu/ql/?c=21980}.
\begin{table}[htbp]
\centering
\small
\setlength{\tabcolsep}{6pt}
\renewcommand{\arraystretch}{1.2}
\begin{tabular}{|l|c|c|c|}
\hline
\textbf{Parameter} & \textbf{O4b} & \textbf{Stage 1} & \textbf{Stage 2} \\
\hline
Power injected & 17 W & 25 W & 40--80 W \\
Signal recycling & Yes & Yes & Yes \\
Squeezing type & FIS & FDS & FDS \\
Squeezing detected level & 0 dB~\footnote{During the commissioning phase in preparation for O4b, the detected squeezing was 2 dB~\cite{VirgoQND}. Following the misalignment of the signal recycling mirror, it dropped to 0 dB.} & 4.5 dB & 5.3 dB \\
Mirror thermal noise @ 100 Hz & $< 6\times10^{-24}/\mathrm{\sqrt{Hz}}$~\cite{VirgoThermal} & $4.8\times10^{-24}/\mathrm{\sqrt{Hz}}$ & $3.4\times10^{-24}/\mathrm{\sqrt{Hz}}$ \\
Technical noise @ 20 Hz & $3\times10^{-22}/\mathrm{\sqrt{Hz}}$ & $(2\times10^{-22}$--$2.6\times10^{-23})/\mathrm{\sqrt{Hz}}$ & $(2\times10^{-22}$--$2.6\times10^{-23})/\mathrm{\sqrt{Hz}}$ \\
BNS range & 55 Mpc & 90--130 Mpc & 120--160 Mpc \\
\hline
\end{tabular}
\caption{Most important parameters used in the calculation of the expected sensitivities for O5. The signal recycling factor and the arm cavity finesse will be the same in O4 and O5.}
\label{tab:virgo_o5_steps}
\end{table}

\section{Conclusions}
The Advanced Virgo Plus upgrade for the O5 observing run represents a significant step forward in gravitational wave detection capabilities. Building upon the lessons learned during O4, the Virgo Collaboration has undertaken a comprehensive redesign of the central interferometer, incorporating stable recycling cavities and enhanced seismic isolation systems. These changes aim to mitigate quantum noise, thermal noise, and technical disturbances, thereby pushing the sensitivity of the detector closer to its theoretical limits.

The implementation of upgrade stages ensures a pragmatic approach to balancing performance goals with resource constraints. Stage 1 upgrades focus on essential improvements (such as the implementation of stable cavities) to ensure reliable operation and increased sensitivity, while Stage 2 upgrades offer additional enhancements that can be pursued as resources allow. With these upgrades, AdV+ is expected to achieve a binary neutron star detection range between 90 and 160 Mpc, with a design sensitivity peak of 180 Mpc, achievable only if technical noise is reduced well below fundamental noise.

This design report outlines the technical roadmap and strategic decisions guiding the AdV+ development for O5. It reflects the Virgo Collaboration’s commitment to advancing gravitational wave astronomy through innovation, precision engineering, and collaborative effort.

\bibliographystyle{unsrt}
\bibliography{sample701}{}

@misc{PhaseOneReport,
  author = {{Virgo Collaboration}},
  title  = {Advanced Virgo Plus Phase I - Design Report},
  year   = {2019},
  note   = {VIR-0596A-19},
  url    = {https://tds.virgo-gw.eu/ql/?c=14430},
}

@misc{PhaseTwoReport,
  author = {{Virgo Collaboration}},
  title  = {Advanced Virgo Plus Phase II - Design Report},
  year   = {2022},
  note   = {VIR-0501C-22},
  url    = {https://tds.virgo-gw.eu/ql/?c=18147},
}

@misc{StabCavReport,
  author = {{Virgo Collaboration}},
  title  = {Short Stable Cavities Preliminary Design},
  year   = {2024},
  note   = {VIR-0461B-24},
  url    = {https://tds.virgo-gw.eu/ql/?c=20521},
}

@misc{matgwinc,
  author = {GWINC},
  title = {matgwinc},
  year = {2020},
  publisher = {GitHub},
  journal = {GitHub repository},
  howpublished = {\url{https://git.ligo.org/gwinc/matgwinc}},
  commit = {f6dd98083242db77317dd5b9411895615fed987d} 
}

@misc{CoatTherm,
  author = {{Davenport}, A. and {Vajente}, G. and {Demos}, N. and {Markosyan}, A. and {Tait}, S. and {Fejer}, M. M. and {Gras}, S. and {Evans}, M. and {Menoni}, C. S.},
  title  = {Minimization of Coating Thermal Noise in TiO2:GeO2 and SiO2 High Reflector Stacks for Gravitational Wave Detectors},
  year   = {2025},
  note   = {LIGO-G2500142},
  url    = {https://dcc.ligo.org/LIGO-G2500142},
}

@misc{LIGOQND,
  author = {{Capote}, E.},
  title  = {Commissioning Report, LLO and LHO},
  year   = {2025},
  note   = {LIGO-G2500585},
  url    = {https://dcc.ligo.org/LIGO-G2500585},
}

@misc{VirgoQND,
  author = {{Vardaro}, M. and {Virgo Collaboration}},
  title  = {Virgo's Quantum Noise Reduction System: Current Status and Future Insights},
  year   = {2024},
  note   = {VIR-0316A-24},
  url    = {https://tds.virgo-gw.eu/ql/?c=20376},
}

@misc{VirgoThermal,
  author = {{Puppo}, P.},
  title  = {Recent investigations on mirror thermal noise in AdV+},
  year   = {2023},
  note   = {VIR-0820A-23},
  url    = {https://tds.virgo-gw.eu/ql/?c=19691},
}

@misc{SSClayout,
  author = {{Flaminio}, R. and {Bonnand}, R.},
  title  = {Short stable recycling cavities: optical lay-out},
  year   = {2024},
  note   = {VIR-1003B-23},
  url    = {https://tds.virgo-gw.eu/ql/?c=19874},
}

@misc{SSCvacta,
  author = {{Lieunard}, B. and {Flaminio}, R. and {Bonnand}, R.},
  title  = {Preliminary design of vacuum tanks for short stable recycling cavities},
  year   = {2023},
  note   = {VIR-1013A-23},
  url    = {https://tds.virgo-gw.eu/ql/?c=19884},
}

@article{AdV_2015,
doi = {10.1088/0264-9381/32/2/024001},
url = {https://doi.org/10.1088/0264-9381/32/2/024001},
year = {2014},
month = {dec},
publisher = {IOP Publishing},
volume = {32},
number = {2},
pages = {024001},
author = {Acernese, F. and et. al.},
title = {Advanced Virgo: a second-generation interferometric gravitational wave detector},
journal = {Classical and Quantum Gravity}
}

@misc{SSCDes,
  author = {{Amar}, W. {Bonnand}, R. and {Flaminio}, R. and {Tournefier} E.},
  title  = {Optical design of stable recycling cavities for the Virgo gravitational wave detector},
  year   = {2026},
  note   = {VIR-0130D-26},
  url    = {https://tds.virgo-gw.eu/ql/?c=22536},
}

@article{Heijningen_2019,
doi = {10.1088/1361-6382/ab075e},
url = {https://doi.org/10.1088/1361-6382/ab075e},
year = {2019},
month = {mar},
publisher = {IOP Publishing},
volume = {36},
number = {7},
pages = {075007},
author = {van Heijningen, J. V. and Bertolini, A. and Hennes, E. and Beker, M. G. and Doets, M. and Bulten, H. J. and Agatsuma, K. and Sekiguchi, T. and van den Brand, J. F. J.},
title = {A multistage vibration isolation system for Advanced Virgo suspended optical benches},
journal = {Classical and Quantum Gravity}
}

@article{InstBaff,
  title = {Performance of an instrumented baffle placed at the entrance of Virgo's end mirror vacuum tower during O5},
  author = {Andr\'es-Carcasona, M. and Mart\'{\i}nez, M. and Mir, Ll. M. and Mundet, J. and Yamamoto, H.},
  journal = {Phys. Rev. D},
  volume = {111},
  issue = {4},
  pages = {042001},
  numpages = {9},
  year = {2025},
  month = {Feb},
  publisher = {American Physical Society},
  doi = {10.1103/PhysRevD.111.042001},
  url = {https://link.aps.org/doi/10.1103/PhysRevD.111.042001}
}

@misc{daumas2026presentation,
author = {Anne Daumas},
title = {Upgrades to the detection optical benches of Virgo in preparation for the next observing run O5},
howpublished = {Oral presentation at Gravitational Waves and Detection Technologies - PAS Rome Meeting 2026},
year = {2026},
address = {Rome, Italy},
url = {https://indico.ego-gw.it/event/915/contributions/9316/},
note = {Accessed: 2026-03-27}
}

@article{Nardecchia_2023,
doi = {10.1088/1361-6382/acb632},
url = {https://doi.org/10.1088/1361-6382/acb632},
year = {2023},
month = {feb},
publisher = {IOP Publishing},
volume = {40},
number = {5},
pages = {055004},
author = {Nardecchia, I. and Minenkov, Y. and Lorenzini, M. and Aiello, L. and Cesarini, E. and Lumaca, D. and Malvezzi, V. and Paoletti, F. and Rocchi, A. and Fafone, V.},
title = {Optimized radius of curvature tuning for the virgo core optics},
journal = {Classical and Quantum Gravity}
}

@article{Accadia_2013,
doi = {10.1088/0264-9381/30/5/055017},
url = {https://doi.org/10.1088/0264-9381/30/5/055017},
year = {2013},
month = {feb},
publisher = {IOP Publishing},
volume = {30},
number = {5},
pages = {055017},
author = {Accadia, T. and et. al.},
title = {Central heating radius of curvature correction (CHRoCC) for use in large scale gravitational wave interferometers},
journal = {Classical and Quantum Gravity}
}

\end{document}